\documentclass[aps,prd,preprint,superscriptaddress,amsmath,amssymb,showpacs]{revtex4-1}
\usepackage{dcolumn}
\usepackage{graphicx}
\usepackage{float}
\usepackage{physics}
\usepackage{xcolor}
\usepackage[colorlinks=true,allcolors=blue]{hyperref}

\usepackage[]{changes}
\usepackage{extarrows}
\renewcommand\sout{\bgroup \color{red} \ULdepth=-.5ex \ULset}

\graphicspath{{figs/}}
\begin{document}
\title{Centrifugal-corrected harmonic oscillator model for spherical proton emitters}

\author{Xiao-Yan Zhu}
%\email[Corresponding author, ]{xyzhu0128@163.com}
\affiliation{School of Mathematics and Physics, University of South China, Hengyang, 421001, China}
\affiliation{School of Nuclear Science and Technology, University of South China, Hengyang, 421001, China}
\author{Wei Gao}
\affiliation{School of Physical Science and Technology, Southwest Jiaotong University, Chengdu, 610031, China}
\author{Jia Liu}%
\affiliation{School of Nuclear Science and Technology, University of South China, Hengyang, 421001, China}
\author{Li-Qiang Zhu}
%\email[Corresponding author, ]{zhuliqiang@mails.ccnu.edu.cn}
\affiliation{Key Laboratory of Quark and Lepton Physics (MOE) and Institute of Particle Physics, Central China Normal University, Wuhan 430079, China} 
\author{Wen-Bin Lin}
%\email[Corresponding author, ]{lwb@usc.edu.cn}
\affiliation{School of Mathematics and Physics, University of South China, Hengyang, 421001, China}
\author{Xiao-Hua Li}
\email[Corresponding author, ]{lixiaohuaphysics@126.com}
\affiliation{School of Nuclear Science and Technology, University of South China, Hengyang, 421001, China}

\date{\today}

\begin{abstract}
In the present work, we propose an improved harmonic oscillator model to systematically evaluate the proton radioactivity half-lives in spherical nuclei, incorporating centrifugal potential effects. By fitting the experimental data, the centrifugal parameter $d = 0.143$ for the correction term $dl(l+1)$ and nuclear potential depth $V_0 = 62.4$ MeV are obtained. The model integrates the relativistic mean field (RMF) theory with the BCS method based on the DD-ME2 force to determine spectroscopic factors $S_p$. Moreover, by verifying the linear relationship between the logarithm of the normalized width $\log_{10}{\gamma^2}$ and fragmentation potential $V_{frag}$, the connection between nuclear structure and tunneling dynamics is confirmed, and an analytical expression for the adjustable parameter $d$ corresponding to the centrifugal potential is derived as $d^{\rm{Ae}}$ $\approx$ 0.167. Compared with $d^{\rm{Ae}}$, the modified model based on $d$ yields results in better agreement with experimental half-lives, and is able to control the error of the experimental data within a factor of 2.4. Furthermore, the extended improved model is used to predict the half-lives of some possible proton radioactivity candidates in NUBASE2020 that are energetically allowed or have been observed but not yet quantified. This work improves the accuracy of proton radioactivity studies and provides a robust theoretical framework for future nuclear structure research.
\end{abstract}

\maketitle
\section{Introduction}
Proton radioactivity is a form of nuclear decay in which unstable atomic nuclei emit protons, resulting in a new nuclide. This process represents the limits of nuclear stability, in which nuclei with an excess of protons spontaneously emit proton to reach a more stable state. 
In 1970, Jackson \emph{et al.} observed the proton transition from the isomeric state of $^{53}\rm{Co}$ to the ground state of $^{52}\rm{Fe}$ \cite{Jackson:1970wid,Cerny:1970zvr},
and the first experimental detection of proton radioactivity from the ground state of $^{151}\rm{Lu}$ was made in 1981 at the GSI velocity filter SHIP \cite{hofmann:1982proton}. 
Subsequent discoveries of proton emissions from $^{147}\rm{Tm}$, $^{109}\rm{I}$, and $^{113}\rm{Cs}$ were made using catcher foil technology in Munich \cite{Faestermann:1984ohr}. In 1984, Hofmann \emph{et al.} detected the decay of isomers in $^{147}\rm{Tm}$ \cite{klepper:1982direct} and $^{150}\rm{Lu}$ \cite{Hofmann:1989}. The development of radioactive beam technology has continuously revealed odd-$Z$ nuclei far from the 
$\beta$-stability line, making them a focal point in nuclear physics research.
To date, about 45 proton emitters with  $51\leq Z\leq 83$ have been discovered, 30 of which are in their ground states, while the others are found in isomeric states \cite{Rudolph:2002zz,Wang:2015loa,Wang:2015loa,Sellin:1993zz,Zhou:2022yzf,Page:1992zza,Manjunatha:2021wyn,Livingston:1993zza}.  
Proton radioactivity not only provides important information about the shell structures \cite{Karny:2008zz,Qi:2012pf} and interactions between bound and unbound states of proton-rich nuclei \cite{Kruppa:2004tf,gamow:1928quantum} but also aids in our understanding of the properties and structure of nuclear matter 
\cite{Delion:2006nh,Zhou:2020vfx,Yuan:2023myz,Zhu:2023akj,Zhu:2022izs,Budaca:2017hkd,Xu:2023etu}, thus becoming a significant area of research in the field of nuclear physics.

Theoretically, proton radioactivity is viewed as a quantum tunneling phenomenon through a potential barrier, and is commonly treated using the Wentzel-Kramers-Brillouin (WKB) approximation. To more accurately describe proton radioactivity and calculate related physical quantities such as half-life, decay energy, orbital angular momentum and preformation factor, also known as spectroscopic factor, various theoretical models have been developed. These include the single folding model \cite{Basu:2005tf}, the Gamow-like model \cite{Liu:2021hmz,Zhang:2023yrn}, the generalized liquid drop model \cite{Dong:2009hr,Wang:2017aym}, the distorted-wave Born approximation \cite{Aberg:1997kfq}, the one-parameter model (OPM) \cite{zou:2022favored}, the density-dependent M3Y (DDM3Y) effective interaction \cite{Bhattacharya:2007sd,qian:2010half}, the Woods-Saxon nuclear potential model \cite{Dudek:1981zz,Alavi:2018cuk,Buck:1992zza}, the phenomenological unified fission model (UFM) \cite{jian:2010unified}, the Coulomb and proximity potential model \cite{Santhosh:2017bzv}, the two-potential approach (TPA) \cite{Qian:2016teg}, the universal decay law (UDL) \cite{Qi:2012pf}, the new Geiger-Nuttall law (N-GNL) \cite{Chen:2019aio}, the phenomenological formula with four-parameter \cite{Sreeja:2018zyq} and so on \cite{Dehghani:2018qfg,ni:2012new,Sreeja:2019lyn,zhang:2009theoretical}. Collectively, these theoretical methods enhance our understanding of proton radioactivity.

Recently, Bayrak \cite{Bayrak:2020ruy} put forward a phenomenological harmonic oscillator model (HOPM) for describing favored $\alpha$ decay half-lives for even–even, even–odd and odd–odd nuclei. Motivated by the shared quantum tunneling mechanism among $\alpha$ decay, cluster and proton radioactivity, we extend the HOPM framework to investigate spherical proton emitters. This extension gives two crucial enhancements: (1) A refined treatment of spectroscopic factor $S_p$ through the RMF method combined with the BCS method using the DD-ME2 interaction, which enables precise nuclear structure characterization. (2) Incorporation of centrifugal potential corrections via a $dl(l+1)$ term to address the heightened sensitivity of proton radioactivity half-lives to orbital angular momentum compared to $\alpha$ and cluster decays. Our improved model demonstrates superior predictive capability, reproducing experimental half-lives within a factor of 2.4 while revealing fundamental structure-emission correlations through derived analytic relationships between reduced widths and fragmentation potentials.

The structure of this article is organized as follows. Section \ref{Sec.II} details the theoretical framework integrating centrifugal corrections with spectroscopic factor calculations. In Section \ref{Sec.III}, the detailed calculations and discussions are presented. Finally, Section \ref{Sec.IV} provides a concise summary.

\section{Theoretical framework}\label{Sec.II}
\subsection{The half-life for spherical proton emission }\label{Sec.II.1}
The half-life of proton radioactivity can be expressed as
\begin{equation}
    T_{1/2}=\frac{\hbar \,ln2}{\Gamma},\label{T1/2}
\end{equation}
where $\hbar$ is the reduced Planck constant. $\Gamma$ represents the proton radioactivity width including the normalized factor $F$ and penetration probability $P$. In semiclassical approximation \cite{Bayrak:2020ruy}, it can be expressed as 
\begin{equation}
   \Gamma=\frac{\hbar^2}{4\,\mu}S_p\, F \,P,\label{Gamma}
\end{equation}
where the reduced mass $\mu=m_pm_d/(m_p+m_d)\approx\ A_d A_p M_{nuc}/(A_d+A_p)$ with $A_d$ and $A_p$ being the mass of the daughter nucleus and emitted proton, respectively. 
$S_p$, the spectroscopic factor of proton emitter, reflects the probability that the orbit occupied by the emitted proton remains unoccupied in the daughter nucleus.

$F$ describes the probability of the emitted proton collision within the inner region, calculated by integrating over this region. It is written as
 \begin{equation}
    F=\frac{1}{\int_{0}^{r_1}\frac{1}{2\,k(r)}dr}.\label{F}
\end{equation}
Under the semiclassical WKB approximation, the barrier penetrability $P$ is given by $P=\exp(-2S)$, where $S$ is the action integral for the proton penetrating the external barrier. It can be expressed as \cite{Bayrak:2020ruy}
 \begin{equation}
 S=\int_{r_1}^{r_2}k(r)\,dr,\label{S-integral}
\end{equation}
where $k(r)=\sqrt{\frac{2\mu}{\hbar^2}|V(r)-Q_p|}$ represents the wave number with $r$ being the separation between the centers of emitted proton and  daughter nucleus. 
$V(r)$ is the total interaction potential of the emitted proton-daughter nucleus. In Eqs. (\ref{F}) and (\ref{S-integral}), the classical turning points are denoted by $r_1$ and $r_2$, which satisfy the conditions $V(r_1)=V(r_2)=Q_p$. Here, $Q_p$ denotes the emitted proton's released energy, it can be given by \cite{zou:2022favored}
\begin{equation}
    Q_p=\Delta M-(\Delta M_d+\Delta M_p)+k(Z^\beta-Z_d^\beta), \label{Qp}
\end{equation}
where the experimental data of the mass excesses $\Delta M$, $\Delta M_d$, and $\Delta M_p$ for the parent nucleus, daughter nucleus, and emitted proton, respectively, are taken from
the most recent atomic mass data NUBASE2020 \cite{Kondev:2021lzi}.
The term $k(Z^\beta-Z_d^\beta)$ represents the screening effect of atomic electrons with $Z_d$ and $Z$ being the proton numbers of daughter and parent nucleus, for $Z \geq 60, k=8.7\rm{eV}, \beta=2.517$ and for $Z < 60, k=13.6\rm{eV}, \beta=2.408$ \cite{Denisov:2005ax}.

During proton emission, the total interaction potential $V(r)$ between the emitted proton and daughter nucleus is typically comprised of the nuclear potential $V_N(r)$, Coulomb potential $V_C$, and centrifugal potential $V_l$, expressed as
\begin{equation}
V(r)=V_N(r)+V_C(r)+V_l(r).
\label{V(r)}
\end{equation}
In this work, we generalize the modified harmonic oscillator potential as the nuclear potential to study proton radioactivity \cite{Bayrak:2020ruy,Delion:2009jw}. It can be written as
\begin{equation}
   V_N(r)=-V_0+V_1\,r^2,
    \label{VN}
\end{equation}
where $V_0$ and $V_1$ represent the depth and diffusivity of nuclear potential, respectively.
In addition, the Coulomb potential $V_C$ is taken as that of a uniformly charged sphere with a sharp radius $R$. It can be expressed as \cite{Buck:1990zz,buck:1991ground}
\begin{eqnarray}
V_C(r) & = & \left\{\begin{array}{ll}
\frac{Z_d e^2}{2R}[3-\frac{r^2}{R^2}], & r\leq r_1, \\
\frac{Z_d e^2}{r}, & r\textgreater r_1,
\end{array}\right.
\end{eqnarray}
where $e^2=1.4399652$ MeV$\cdot$fm represents the square of the elementary charge of an electron, and $R$ denotes the sharp radius calculated by the semi-empirical formula as
\begin{equation}
   R=r_0A_d^{1/3}+R_p.
    \label{R}
\end{equation}
Here, $A_d$ is the mass number of the daughter nucleus. $R_p$ is the matter radius of the proton. In this study, we choose $R_p=0.8409$ fm and $r_0=1.14$ fm \cite{Chen:2019jnc}. For the centrifugal potential $V_l$, it can be written as
\begin{equation}
   V_l(r)=\frac{\hbar^2\,l(l+1)}{2\mu r^2},
    \label{Vl}
\end{equation}
where $l$ is the orbital angular momentum carried away by the emitted proton, and the minimum angular momentum $l_{min}$ can be obtained by the conservation laws of spin and parity. 

\begin{figure}[!htb]\centering
\includegraphics
  [width=0.75\hsize]
  {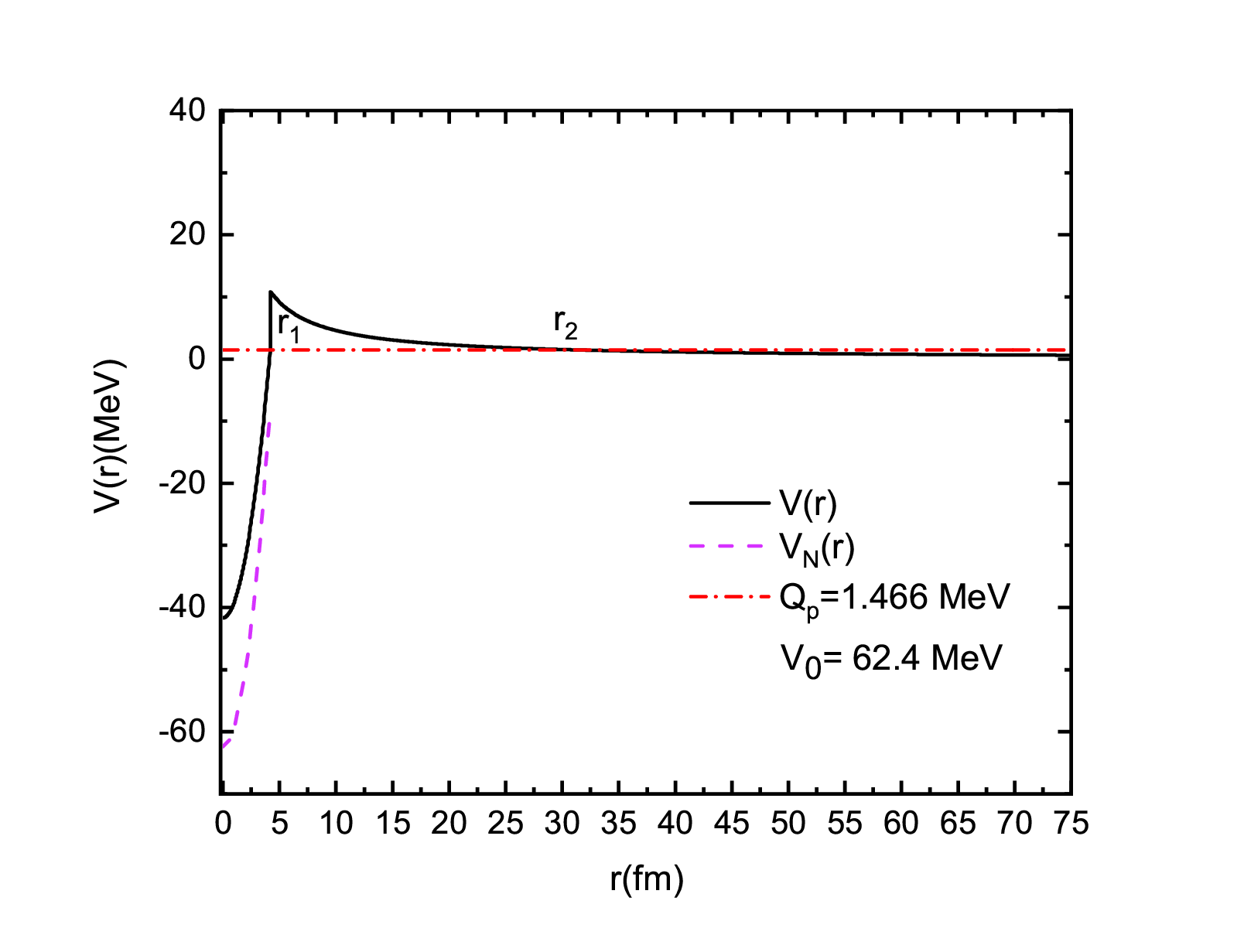}
\caption{(color online) The schematic diagram of the total interaction potential $V(r)$ and the modified harmonic oscillator potential $V_N(r)$ versus $r$.}
\label{Vfinal}
\end{figure}

As for favored proton radioactivity, the total interaction potential $V(r)$ between the emitted proton and daughter nucleus can be written as
\begin{eqnarray}
V(r) & = & \left\{\begin{array}{ll}
 C_0-V_0+(V_1-C_1)r^2, & r\leq r_1, \\
 \frac{C_2}{r}, & r\textgreater r_1,
\end{array}\right.
\label{Vr}
\end{eqnarray}
where $C_0=\frac{3Z_d\,e^2}{2R}$, $C_1=\frac{Z_d\,e^2}{2R^3}$ and $C_2=Z_de^2$. The modified harmonic oscillator potential $V_N(r)$ is explicitly truncated at the classical turning point $r_1$, defined by $V(r_1)=Q_p$. For $r\leq r_1$, both nuclear and Coulomb terms contribute to $V(r)$. For $r>r_1$, $V_N(r)$ vanishes completely, leaving only the Coulomb potential $V_C$. 
As an example to intuitively describe $V(r)$ and $V_N(r)$, the nucleus $^{155}\rm{Ta}$ in Fig. \ref{Vfinal} is shown. Based on the conditions $V(r_1)=V(r_2)=Q_p$, the values of $r_1$ and $r_2$ are derived as $r_1=\sqrt{\frac{Q_p+V_0-C_0}{V_1-C_1}}$ and $r_2=\frac{C_2}{Q_p}$, respectively. 

The Bohr-Sommerfeld quantization condition is regarded as an essential part of determining the quantum state in the WKB approximation \cite{Kelkar:2007wy,Chen:2021tmc}. In this work, the analytical expression for the nuclear potential diffusivity $V_1$ is obtained by applying this condition, which is given by
\begin{equation}
\int_{0}^{r_1}\sqrt{\frac{2\mu}{\hbar^2}(V(r)-Q_p)}dr=(G-l+1)\frac{\pi}{2},\label{Bsq}
\end{equation}
where $G=2n_r+l$ represents the global principal quantum number with $n_r$ and $l$ being the radial quantum number and the angular momentum quantum number, respectively. For proton radioactivity, we select $G=4$ or $5$, corresponding to the $4\hbar\omega$ or $5\hbar\omega$ oscillator shells of the emitted proton. The relationship between $V_0$ and $V_1$ can be analytically derived by using the above Eq. (\ref{Bsq}). It is written as
\begin{equation}
   V_1=C_1+\frac{\mu}{2\hbar^2}\bigg(\frac{Q_p+V_0-C_0}{1+G}\bigg)^2,\label{V1}
\end{equation}
with the conditions $C_0\textless(Q_p+V_0)$ and $C_1\textless V_1$ need to be satisfied.
On the basis of Eq. (\ref{V1}), the normalization factor $F$ in Eq. (\ref{F}) and the action integral $S$ in Eq. (\ref{S-integral}) can be further analytically written as
\begin{equation}
    F=\frac{4}{\pi}\,\frac{\mu}{\hbar^2}\bigg(\frac{Q_p+V_0-C_0}{1+G}\bigg),\label{FF}
\end{equation}
\small
\begin{equation}
 S=\frac{\sqrt{2\mu}}{\hbar}\frac{C_2}{\sqrt{Q_p}}\bigg (\rm{arccos}\bigg (\sqrt{\frac{Q_pr_1}{C_2}}\bigg)\!-\!\sqrt{\frac{Q_pr_1}{C_2}\!-\!\bigg (\frac{Q_pr_1}{C_2}\bigg)^2}\bigg).\label{SS}
\end{equation}
\small
Consequently, the logarithmic form of the half-life for favored proton radioactivity can be expressed as
\begin{equation}
    {\rm{log}}_{10}T_{1/2}= {\rm{log}}_{10}\bigg( \frac{\pi\, \hbar \,ln2}{S_p} \frac{1+G}{Q_p+V_0-C_0}\bigg)+2S{\rm{log}}_{10}(e).\label{logT}
\end{equation}

In the case of favored proton radioactivity, should the orbital angular momentum of the proton being emitted be zero, the influence exerted by the centrifugal potential equally becomes null. Nevertheless, during spherical proton emission, the centrifugal potential due to an orbital angular momentum ($l\neq 0$) raises the potential barrier height, thereby affecting the penetration probability and the associated decay process. 
Compared to $\alpha$ decay and cluster radioactivity, the half-life of proton radioactivity exhibits a greater sensitivity to both the decay energy $Q_p$ and orbital angular momentum $l$. Hence, it is necessary to consider the impact of the centrifugal potential on spherical proton emission.
In this work, the term $dl(l+1)$ \cite{Soylu:2021tmy,Zhu:2024swx} is introduced to consider the effect of centrifugal potential on the spherical proton radioactivity in Eq. (\ref{logT}), in a manner similar to the method used in the study of $\alpha$ decay \cite{Qian:2012zza}. Therefore, an improved model for the half-life of a spherical proton emitter is expressed as
\begin{equation}
    {\rm{log}}_{10}T_{1/2}= {\rm{log}}_{10}\bigg( \frac{\pi\, \hbar \,ln2}{S_p}\frac{1+G}{Q_p+V_0-C_0}\bigg)+2S{\rm{log}}_{10}(e)+ d\,l\,(l+1).\label{logTT}
\end{equation}

\subsection{The spectroscopic factor of proton radioactivity}\label{Sec.II.2}
It is assumed that the core nucleus remains unaltered throughout the decay process. In this study, the spectroscopic factor $S_p$ for the proton-daughter system is calculated using the RMF theory combined with the BCS method \cite{Dong:2009hr,Qian:2016teg}.
 The RMF theory is particularly well suited to investigating the single particle structure of rich proton nuclei based on the Dirac-Lagrangian density, as it naturally incorporates the spin degree of freedom \cite{Bhattacharya:2007sd,qian:2010half}. It can be estimated by 
\begin{equation}
    S_{p}^{\rm{{Cal}}}=u_j^{2},\label{Spcal}
\end{equation}
where $u_j^{2}$ represents the probability that the orbit of the emitted proton is empty in the daughter nucleus. In this work, nuclear pairing correlations are treated using the BCS method, and the pairing gaps for protons and neutrons are expressed as functions of the mass number A, i.e., $\Delta n = \Delta p=11.2\,A^{-1/2}$MeV \cite{Qian:2016teg}.

\subsection{Analytic relation between fragmentation potential and reduced width }\label{Sec.III.3}
The nuclear potential in Eq. (\ref{Vr}) can be rigorously mapped to the shifted harmonic oscillator (HO) potential in Ref. \cite{Delion:2009jw}. It is taken that 
\begin{equation}
    \beta=\frac{2V_1}{\hbar\omega },\label{B}
\end{equation}
and
\begin{equation}
   v_0=-V_0+\frac{1}{2}\hbar\omega\beta r_0^2,
\end{equation}
our potential becomes identical to Eq. (3.1) of Ref. \cite{Delion:2009jw} when $r_0=0$. The continuity condition at $r_B$ is 
\begin{equation}
   V_1r_B^2-V_0=\frac{Z_de^2}{r_B}-Q_p\equiv V_{frag},\label{V_11}
\end{equation}
which establishes the fragmentation potential $V_{frag}$ as the energy difference between Coulomb barrier and $Q_p$.

For $r\leq r_B$, the part under the integral in Eq. (\ref{Bsq}) is changed by
\begin{equation}
V(r)-Q_p=-V_0+V_1r^2-Q_p=V_1\bigg(r^2-\frac{V_0+Q_p}{V_1}\bigg).
\end{equation} 
The harmonic oscillator eigenstate condition is introduced, where $Q_p$ corresponds to the ground-state energy of the harmonic oscillator, that is $Q_p=\mu_0+\frac{3}{2}\hbar\omega$. However, according to Eq. (3.2) of Ref. \cite{Delion:2009jw}, the shifted harmonic oscillator is approximated $Q_p \approx \mu_0+\frac{1}{2}\hbar\omega$. Then the product differentiates the simple Eq. (\ref{Bsq}), which is given by
\begin{equation}
\int_{0}^{r_1}\sqrt{\frac{2\mu V_1}{\hbar^2}(r^2-a^2)}dr=(G-l+1)\frac{\pi}{2},
\end{equation}
where $a^2=(V_0+Q_p)/V_1$. For ground-state transitions, $V_1$ is rewritten as
\begin{equation}
V_1=\frac{\mu}{2\hbar^2}\bigg[\frac{\pi \hbar(1+G)}{2r_1^2}\bigg]^2. \label{V_111}
\end{equation}
It is important to note that Eq. (\ref{V_111}) is analytically derived under the ground-state \cite{Delion:2009jw}. For excited state transitions, the action integral in Eq. (\ref{Bsq}) is evaluated with the global quantum number $G$ adapted to their dominant oscillator shells, while retaining the same functional form of $V_1$. This ensures consistent treatment of centrifugal corrections across both ground and excited states within the harmonic oscillator approximation.

Following the methodology of Delion \cite{Delion:2009jw} and to ensure full consistency within our harmonic oscillator framework for $l\neq0$ decays, we employ the three-dimensional isotropic harmonic oscillator radial wave function. For a state characterized by radial node number $n_r$ and orbital angular momentum $l$, the probability density at the barrier radius $r_{B}$ is proportional to
\begin{equation}
\left|f_0^{(int)}\left(r_B\right)\right|^2 \propto (r_B/b)^{2l}e^{-r_B^2/b^2}[L_{n_r}^{l+1/2}(r_B^2/b^2)]^2,\label{f0}
\end{equation}
where $b=\sqrt{\hbar/(\mu\omega)}$ is the oscillator length and $L_{n_r}^{l+1/2}$ is an associated Laguerre polynomial. Substituting the relation $r_B^2=(V_{frag}+V_{0})/V_{1}$ from Eq. (\ref{V_11}) and using Eq. (\ref{B}), the exponent becomes $-2V_{1}r_B^2/(\hbar\omega)=-2(V_{frag}+V_0)/(\hbar\omega)$. Therefore, the essential dependence of the internal wave function amplitude on $V_{frag}$ is preserved as
\begin{equation}
\left|f_0^{(int)}\left(r_B\right)\right|^2 = A_l^2 \exp\left(-\frac{2V_1}{\hbar \omega}r_B^2\right)=A_l^2 \exp\left(-\frac{2(V_{frag}+V_0)}{\hbar \omega}\right),\label{f0}
\end{equation}
where $A_l^2$ is the $l$-dependent pre-exponential factors and normalization. Substituting Eqs. (\ref{B}) and (\ref{V_11}) to Eq. (\ref{f0}), and then combined with Eq. (2.2) in Ref. \cite{Delion:2009jw}, we obtain the following linear relationship
\begin{equation}
\log_{10}{\gamma^2} =-\frac{2\log_{10}e}{\hbar \omega}V_{frag}+\log_{10}\left(\frac{\hbar^2A_l^2}{2\mu r_{B}}\right)-\frac{2V_0log_{10}e}{\hbar \omega},\label{f00}
\end{equation}
where the slope $-\frac{2\log_{10}e}{\hbar \omega}$ encodes nuclear stiffness, reflecting sensitivity to structural changes. The intercept term $\log_{10}\left(\frac{\hbar^2A_l^2}{2\mu r_{B}}\right)$ depends on wave-function normalization and reduced mass. Eq. (\ref{f00}) analytically establishes that the reduced width $\gamma^2$ decreases linearly with the fragmentation potential $V_{frag}$. Since $\gamma^2$ quantifies the probability amplitude of proton radioactivity at the barrier radius, a larger $V_{frag}$ (higher Coulomb barrier relative to $Q_p$) suppresses $\gamma^2$, thereby prolonging the half-life $T_{1/2}$. This chain $V_{frag} \rightarrow \gamma^2 \rightarrow T_{1/2}$ unifies nuclear structure effects with tunneling dynamics in proton emission.

The contribution of the centrifugal potential in Eq. (\ref{Vl}) to the fragmentation potential $V_{frag}$ at the barrier radius $r_B$ is 
\begin{equation}
V_{frag}'=V_{frag}+\frac{\hbar^2\,l(l+1)}{2\mu r^2}.\label{Vfragg}
\end{equation}
Finally, combining the equations (\ref{logTT}) and (\ref{f00}), the analytic expression $d^{Ae}$ of the adjustable parameter $d$ is expressed as
\begin{equation}
d^{\rm{Ae}}=\frac{\hbar}{\mu r_1^2 \omega \rm{ln10}}.\label{d}
\end{equation}

\section{Results and discussion}\label{Sec.III}

In our previous study \cite{Zhu:2023izb}, we systematically described the cluster radioactivity half-lives by accounting for preformation probabilities based on the HOPM. Given that cluster radioactivity shares the same mechanism as proton radioactivity, we attempt to generalize this model to investigate the half-lives of spherical proton emissions. It is worth noting that the proton radioactivity half-life is more sensitive to orbital angular momentum than cluster radioactivity \cite{Sonzogni:2002slb}. As a consequence, in this work, considering the spectroscopic factor $S_{p}$ and the effect of the centrifugal potential, we propose an improved model to evaluate the proton radioactivity half-lives for spherical nuclei. 

The spectroscopic factor $S_{p}$, which involves a variety of nuclear structure properties, is also called the formation probability and is crucial in half-life calculation.
Some semi-microscopic and phenomenological methods are used to calculate the spectroscopic factor of proton radioactivity 
\cite{Aberg:1997kfq,qian:2010half,Rong:2021bim,Li:2013oza,Delion:2009jw,delion:2006theories}.
In this work, the spectroscopic factor $S_{p}$ is obtained by using the RMF theory and BCS method with the force parameter chosen as DD-ME2 in Eq. (\ref{Spcal}), which has demonstrated widespread success and applicability in depicting diverse structural characteristics across a broad spectrum of nuclei \cite{Bhattacharya:2007sd,Dong:2009hr,Qian:2016teg,Rong:2021bim,Li:2013oza}. 
The spectroscopic factor for spherical nuclei is given in the fifth column of Table \ref{table1}, denoted by $S_{p}^{\rm{{cal}}}$. 

\begin{figure}[!htb]\centering
 \includegraphics
  [width=0.85\hsize]
  {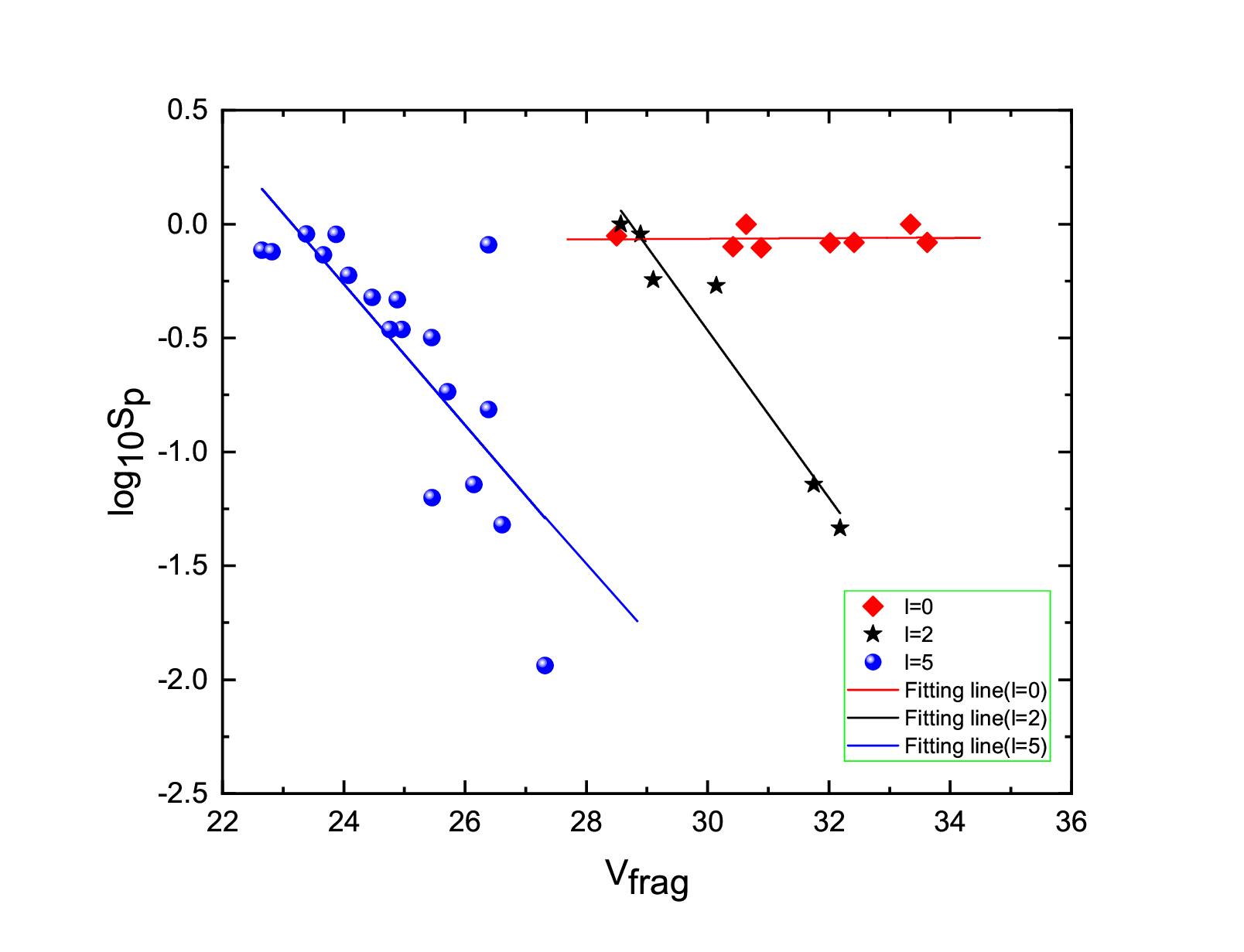}
 \caption{(color online) The linear relationship between the logarithm of spectroscopic factors $S_{p}^{\rm{{cal}}}$ calculated by Eq. (\ref{Spcal}) and the fragmentation potential $V_{frag}$.}
 \label{logSp-Vfrag}
 \end{figure}

Delion \cite{Delion:2009jw} recently introduced a universal formula that connects the logarithm of the squared reduced width to the fragmentation potential $V_{frag}$. It was found that the relationship between the spectroscopic factor and the mass of the emitted proton can be well explained by the fragmentation potential $V_{frag}$, which is given by the difference between the Coulomb barrier $V_c$ and released energy $Q_p$. It can be written as
\begin{equation}
    V_{frag}=\frac{Z_d\,e^2}{r_1}-Q_p.\label{Vfrag}
\end{equation}
As a means of verification, Fig. \ref{logSp-Vfrag} shows the piecewise linear relationship between $\rm{log}_{10}S_p$ and $V_{frag}$ with distinct $l$ corresponding to different slopes, reflecting the influence of the centrifugal potential on the amplitude of wave functions. 
The correlation in Fig. 2 stems from the sensitivity of RMF+BCS-calculated spectroscopic factors to nuclear potential structure. Since $S_p \propto u_j^{2}$, and $u_j^{2}$ is associated with the amplitude of single-particle wave functions near the Fermi level, variations in the fragmentation potential indirectly reflect differences in occupation probabilities of proton emission orbitals for different $l$ values.
For $l\neq 0$ emissions, centrifugal potential modifications cause the wave function amplitude $u_j^{2}$ near the Fermi level to deviate from 0.5, leading to dispersion in $S_p$. Such as when $l=5$, the high angular momentum significantly alters the barrier penetration conditions, resulting in a smaller $u_j^{2}$ ($S_p \sim 0.1–0.3$  in Table \ref{table1}), which contrasts with the $l=0$ case ($S_p \sim 0.7-0.9$). 
This further substantiates that the spectroscopic factor of proton radioactivity can be reasonably described by the RMF theory combined with the BCS method in this work.

\begin{figure}[!htb]\centering
 \includegraphics
  [width=0.75\hsize]
  {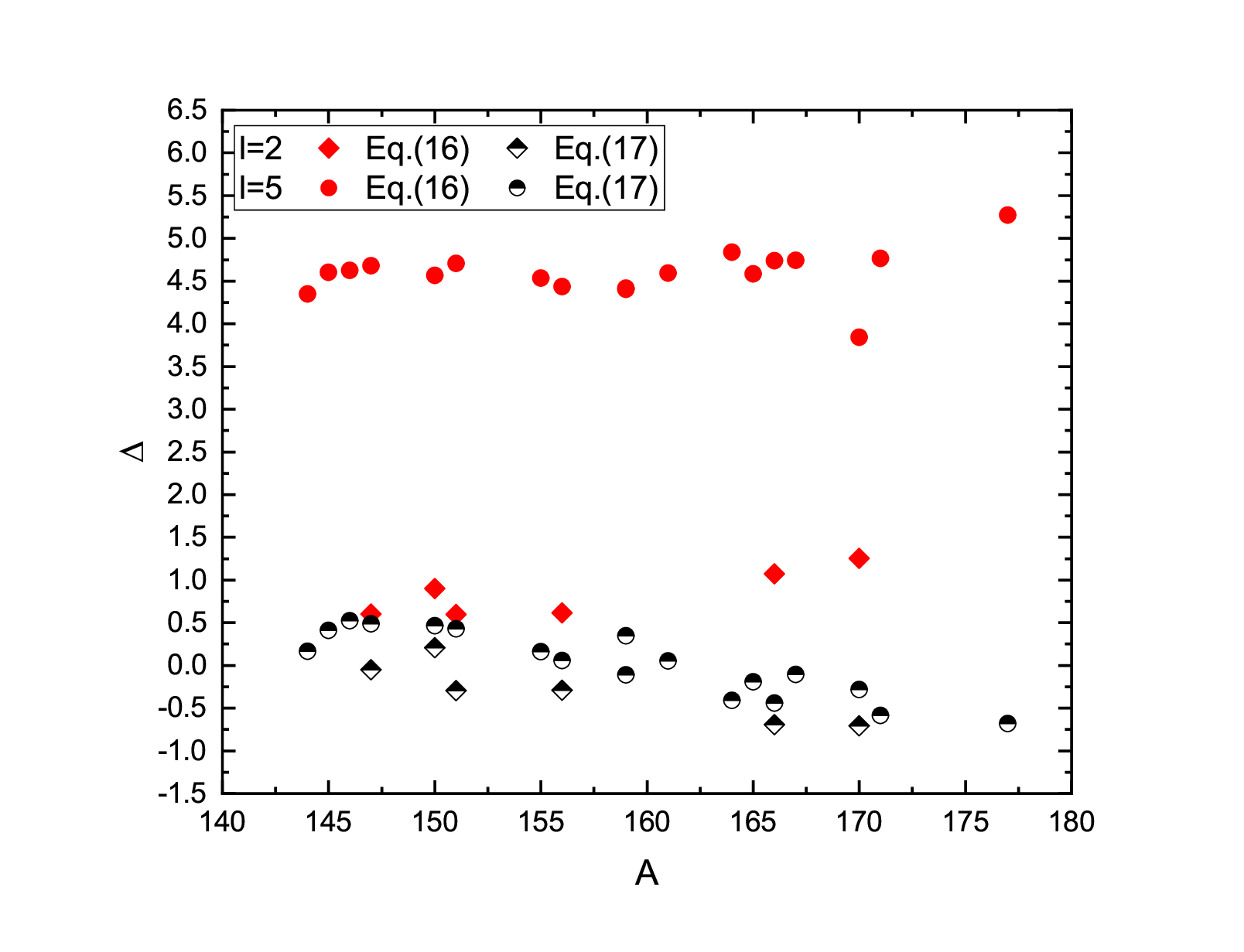}
 \caption{(color online) The logarithmic differences between experimental half-lives and calculate values for unfavored proton radioactivity. The different colors represent the different angular momentum taken away by the proton emitters. For each angular momentum cases, the squares and circles are represented by Eq. (\ref{logT}), while the pentagrams and triangles are denoted by Eq. (\ref{logTT}), respectively.}
 \label{wuchal25}
 \end{figure}

With the above confirmation of the reliability of the spectroscopic factors, based on the $S_{p}$ values obtained from Eq. (\ref{Spcal}), we determine the adjustable parameter $V_0=62.4$ MeV in Eq. (\ref{logT}) by fitting the experimental half-lives of favored proton radioactivity.
For spherical proton emissions, we directly introduce the term $dl(l+1)$ in Eq. (\ref{logT}) to account for this effect of the centrifugal potential, as shown in Eq.(\ref{logTT}). By fitting experimental proton radioactivity half-lives of spherical nuclei, we obtain the adjustable parameter $d$ in Eq. (\ref{logTT}) as $d=0.143$. Based on the obtained adjustable parameters $V_0$ and $d$, the differences $\Delta$ between the experimental data of proton radioactivity half-lives and the calculated values using Eq. (\ref{logTT}) for 32 spherical nuclei in logarithmic form are plotted in Fig. \ref{wuchal25}, denoted as black symbols. The red symbols in this figure represent the differences $\Delta$ obtained from Eq. (\ref{logT}). As can be seen from the figure, the $\Delta$ between the experimental data and calculated values using Eq. (\ref{logT}) for some nuclei becomes large. In particular, when $l=5$, the calculated values are nearly 6 orders of magnitude smaller than experimental data. Therefore, the effect of the centrifugal potential on the proton radioactivity cannot be ignored as $l$ increases. The aim of this work is also to generalize the HOPM Eq. (\ref{logT}) to spherical proton emission and to propose an improved model for proton radioactivity half-lives. The differences $\Delta$ between the experimental proton radioactivity half-lives and the calculated values in logarithmic form can be expressed as
\begin{equation}
    \Delta={\rm{log}}_{10}T_{1/2}^{\rm{{Exp}}}-{\rm{log}}_{10}T_{1/2}^{\rm{{Cal}}},\label{Delta}
\end{equation}
where ${\rm{log}}_{10}T_{1/2}^{\rm{{Exp}}}$ and ${\rm{log}}_{10}T_{1/2}^{\rm{{Cal}}}$ denote the logarithm of the experimental half-life and calculated values, respectively.

\begin{figure}[!htb]\centering
 \includegraphics
  [width=0.75\hsize]
  {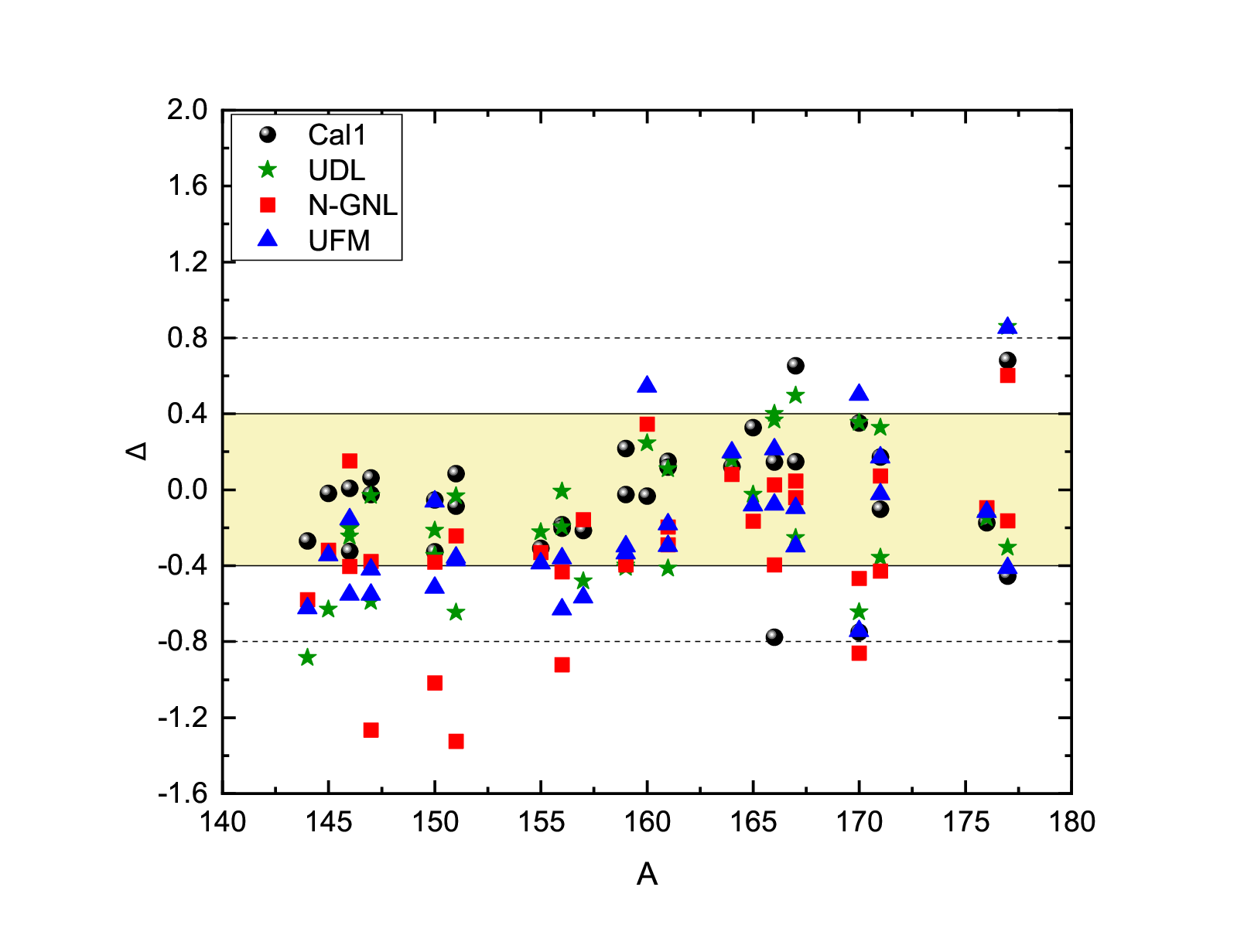}
 \caption{(color online) Deviations between the theoretical proton radioactivity half-lives and the experimental values.}
 \label{wucha}
 \end{figure}

In the following, based on the obtained parameters $V_0=62.4$ MeV and $d=0.143$, we systematically calculate the proton radioactivity half-lives of spherical nuclei by using Eq. (\ref{logTT}) with the spectroscopic factors taken from Eq. (\ref{Spcal}). 
For comparison, UDL \cite{Qi:2012pf}, N-GNL \cite{Chen:2019aio}, UFM \cite{jian:2010unified} are also used. The detailed calculations are listed in Table \ref{table1}. In this table, the first four columns provide the proton emission, the released energy $Q_p$, the spin and parity transition $(j_p^{\pi }\rightarrow j_d^{\pi})$, and the angular momentum $l$ carried away by the emitted proton, respectively. The fifth column lists the calculated spectroscopic factor $S_{p}^{\rm{{Cal}}}$ by Eq. (\ref{Spcal}). 
The seventh and eighth columns as well as the tenth to twelfth columns give the logarithmic form of the experimental proton emissions half-lives and calculated values using Eq. (\ref{logTT}) with $S_p$ obtained by Eq. (\ref{Spcal}), UDL \cite{Qi:2012pf}, N-GNL \cite{Chen:2019aio}, and UFM \cite{jian:2010unified}, which are expressed as ${\rm{log}}_{10}T_{1/2}^{\rm{{Exp}}}$,
${\rm{log}}_{10}T_{1/2}^{\rm{{Cal1}}}$, ${\rm{log}}_{10}T_{1/2}^{\rm{{UDL}}}$, ${\rm{log}}_{10}T_{1/2}^{\rm{{N-GNL}}}$ and ${\rm{log}}_{10}T_{1/2}^{\rm{{UFM}}}$, respectively. 
As can be seen from Table \ref{table1}, compared to other results, the calculated proton radioactivity half-lives with the obtained spectroscopic factors by Eq. (\ref{Spcal}) can better reproduce the experimental data except for a few nuclei such as $ ^{166}\rm{Ir^{m}}$, $ ^{170}\rm{Au}$ and $ ^{177}\rm{Tl^{m}}$. It is found that the spectroscopic factors $S_p$ are quite small when the daughter nuclei are close to the proton layer. This may be due to differences in the selection of pairing energy gaps and the number of basis states used in the calculation.
For clearer visualization, the differences $\Delta$ between the experimental half-lives of proton radioactivity and calculated values, ${\rm{log}}_{10}T_{1/2}^{\rm{{Cal1}}}$, 
${\rm{log}}_{10}T_{1/2}^{\rm{{UDL}}}$, ${\rm{log}}_{10}T_{1/2}^{\rm{{N-GNL}}}$ and ${\rm{log}}_{10}T_{1/2}^{\rm{{UFM}}}$, are shown in Fig. \ref{wucha} respectively denoted as the black balls, green pentagrams, red squares and blue upper triangles. 
From this figure, one can clearly see that our model (Cal1) predominantly clusters within [-0.4, 0.4], whereas UDL and N-GNL exhibit greater dispersion, exemplified by $^{147}\rm{Tm}$ with $\Delta_{\rm{UDL}}=0.732$ versus $\Delta_{\rm{Cal1}}=0.187$. Notably, the centrifugal correction demonstrates critical efficacy for high-angular momentum nucleus: For $ ^{155}\rm{Ta}$ ($l$=5), $\Delta_{\rm{Cal1}}=-0.16$ shows marked improvement over $\Delta_{\rm{UDL}}=-0.223$ and $\Delta_{\rm{UFM}}=-0.710$. Meanwhile, it also explains the feasibility of obtaining $S_p$ for spherical proton radioactivity using Eq. (\ref{Spcal}) and verifies the reliability of the improved model Eq. (\ref{logTT}) by considering the effect of centrifugal potential.

\begin{table*}[!htbp]
    \centering
    \setlength\tabcolsep{0.8pt}
    \renewcommand\arraystretch{0.75}
    \caption{Comparison between the experimental and calculated proton radioactivity half-lives for spherical nuclei.  The symbol $'\#'$ represents estimated values based on trends in neighboring nuclides with the same $Z$ and $N$ parities. The symbol $m$ denotes the isomeric state, and $'()'$ denotes uncertain spin and/or parity, the experimental data for proton radioactivity half-lives, $Q_p$ values and spin-parity information are obtained from Refs. \cite{Xu:2023etu,Zhang:2023yrn}}
	\begin{tabular}{cccccccccccc}	\hline
        \multicolumn{3}{c}{}  &  \multicolumn{9}{c}{$\log_{10}{T_{1/2 } (s)} $} \\\cline{7-12}
       Nuclei & $Q_{p}$(MeV) & $j_{p}^{x}\longrightarrow j_{d}^{x}$ & $l$\,\,\,\,\, &\,\,\,$S_p^{\rm{Cal}}$\,\,\,&\,\,\,$d^{\rm{Ae}}$ \,\,\,&\,\,\, Exp \,\,\, & \,\,\,Cal1\,\,\,\,\,& \,\,\,\,\,Cal2\,\,\, &\,\,\, UDL &\, N-GNL\,\,&UFM \\\hline
        $ ^{144}\rm{Tm}$ & 1.724 & $(10^{+})\rightarrow 9/2^{-}\#$     &5\,\,  &0.769&0.146&-5.569 & -5.734&-5.994 & -4.685 & -4.989  & -4.832\\
        $ ^{145}\rm{Tm}$ & 1.754 & $(11/2^{-})\rightarrow 0^{+}$       &5\,\,  &0.756&0.146&-5.499 & -5.908&-6.158 & -4.869 & -5.180 & -5.033\\
        $ ^{146}\rm{Tm}$ & 0.904 & $(1^{+})\rightarrow (1/2^{+})$      &0\,\,  &0.889&0.205&-0.810 & -0.975&-1.305 & -0.605 & -0.963  & -0.603\\
        $ ^{146}\rm{Tm^{m}}$ & 1.214 & $(5^{-})\rightarrow (1/2^{+})$  &5\,\,  &0.906&0.144&-1.137 & -1.660&-1.987 & -0.893 & -0.733  & -0.542 \\
        $ ^{147}\rm{Tm^{m}} $& 1.133 & $3/2^{+}\rightarrow 0^{+}$      &2\,\,  &0.999&0.207&-3.444 & -3.393&-3.333 & -2.855 & -2.179  & -3.023\\
        $ ^{147}\rm{Tm} $& 1.072 & $11/2^{-}\rightarrow 0^{+}$         &5\,\,  &0.732&0.144&0.587  & 0.100 &-0.244 & 0.618  & 0.965   & 1.274\\
        $ ^{150}\rm{Lu^{m}} $&1.305 & $(1+,2^{+})\rightarrow (1/2^{+})$&2\,\,  &0.901&0.205&-4.398 & -4.605&-4.563 & -4.050 & -3.381  & -4.292\\
        $ ^{150}\rm{Lu}$& 1.285 & $(5^{-})\rightarrow (1/2^{+})$       &5\,\,  &0.901&0.142&-1.347 &-1.812 &-2.196 & -1.132 & -0.965  & -0.785\\
        $ ^{151}\rm{Lu^{m}}$ & 1.315 & $3/2^{+}\rightarrow 0^{+}$      &2\,\,  &0.569&0.205&-4.796 & -4.501&-4.457 & -4.150 & -3.472  & -4.203 \\
        $ ^{151}\rm{Lu}$ & 1.255 & $11/2^{-}\rightarrow 0^{+}$         &5\,\,  &0.597&0.142&-0.896 & -1.322&-1.705 & -0.863 & -0.654  & -0.300\\
        $ ^{155}\rm{Ta}$& 1.466 & $11/2^{-}\rightarrow 0^{+}$          &5\,\,  &0.477&0.141&-2.495 & -2.655&-3.073 & -2.272 & -2.164  & -1.785\\
        $ ^{156}\rm{Ta} $&1.036 & $(2^{-})\rightarrow 7/2^{-}\#$       &2\,\,  &0.536&0.201&-0.826 & -0.537&-0.535 & -0.630 & 0.096   & -0.193 \\
        $ ^{156}\rm{Ta^{m}}$ & 1.126  & $(9^{+})\rightarrow 7/2^{-}\#$ &5\,\,  &0.466&0.140&0.933  & 0.875 &0.410 & 0.942  & 1.366   & -1.896\\
        $ ^{157}\rm{Ta}$ & 0.946 & $1/2^{+}\rightarrow 0^{+}$          &0\,\,  &0.795&0.201&-0.527 & -0.171&-0.522 & -0.045 & -0.369  & -0.140\\
        $ ^{159}\rm{Re^{m}}$ & 1.816 & $11/2^{-}\rightarrow 0^{+}$     &5\,\,  &0.344&0.144&-4.665 & -5.013&-5.354 & -4.267 & -4.282  & -4.366\\
        $ ^{159}\rm{Re}$ & 1.816 & $11/2^{-}\rightarrow 0^{+}$         &5\,\,  &0.344&0.141&-4.678 & -4.569&-4.999 & -4.268 & -4.284  & -3.904\\
        $ ^{160}\rm{Re}$ & 1.267 & $(4^{-})\rightarrow 7/2^{-}\#$      &0\,\,  &0.997&0.200&-3.163 &-3.928 &-4.282 & -3.408 & -3.508  & -3.704\\
        $ ^{161}\rm{Re}$ & 1.216& $1/2^{+}\rightarrow 0^{+}$           &0\,\,  &0.786&0.200&-3.306 & -3.247&-3.602 & -2.893 & -3.017  & -3.020 \\
        $ ^{161}\rm{Re^{m}}$ & 1.336 & $11/2^{-}\rightarrow 0^{+}$     &5\,\,  &0.317&0.139&-0.678 &-0.731 &-1.225 & -0.788 & -0.481  & -0.117\\
        $ ^{164}\rm{Ir}$ & 1.844 & $(9^{+})\rightarrow 7/2^{-}$        &5\,\,  &0.063&0.140&-3.959 &-3.552 &-4.039 & -4.114 & -4.039  & -2.956\\
        $ ^{165}\rm{Ir^{m}}$ & 1.727 & $(11/2^{-})\rightarrow 0^{+}$   &5\,\,  &0.184&0.139&-3.433 &-3.241 &-3.741 & -3.409 & -3.267  & -2.615\\
        $ ^{166}\rm{Ir}$ & 1.167 & $(2^{-})\rightarrow (7/2^{-})$      &2\,\,  &0.072&0.198&-0.824 &-0.131 &-0.172 & -1.189 & -0.428  & -0.105\\
        $ ^{166}\rm{Ir^{m}}$ & 1.347 & $(9^{+})\rightarrow(7/2^{-})$   &5\,\,  &0.072&0.138&-0.076 & -0.363&-0.190 & -0.477 & -0.102  & -1.145\\
        $ ^{167}\rm{Ir}$ & 1.087 & $1/2^{+}\rightarrow0^{+}$           &0\,\,  &0.828&0.197&-1.120 &-0.972 &-1.342 & -0.867 & -1.078  & -0.739\\
        $ ^{167}\rm{Ir^{m}}$ & 1.262 & $11/2^{-}\rightarrow 0^{+}$     &5\,\,  &0.153&0.138& 0.842 & 0.949 &0.388  & -0.346 & 0.796   & -1.753\\
        $ ^{170}\rm{Au}$ & 1.487 & $(2^{-})\rightarrow (7/2^{-})$      &2\,\,  &0.046&0.197&-3.487 &-2.779 &-2.827 & -3.841 & -3.020  & -2.650\\
        $ ^{170}\rm{Au^{m}}$ & 1.767 & $(9^{+})\rightarrow (7/2^{-})$  &5\,\,  &0.811&0.138&-3.975 &-3.691 &-4.245 & -3.331 & -3.115  & -3.140\\
        $ ^{171}\rm{Au}$ & 1.464 & $1/2^{+}\rightarrow 0^{+}$          &0\,\,  &0.831&0.197&-4.652 &-4.681 &-5.052 & -4.294 & -4.224  & -4.549\\
        $ ^{171}\rm{Au^{m}}$ & 1.702 & $11/2^{-}\rightarrow 0^{+}$     &5\,\,  &0.048&0.138&-2.587 &-2.000 &-2.558 & -2.915 & -2.660  & -1.439\\
        $ ^{176}\rm{Tl}$ & 1.278 &$(3^{-},4^{-})\rightarrow (7/2^{-})$ &0\,\,  &0.999&0.194&-2.208 &-2.254 &-2.642 & -2.059 & -2.113  & -2.091\\
        $ ^{177}\rm{Tl}$ & 1.173 & $(1/2^{+})\rightarrow 0^{+}$        &0\,\,  &0.832&0.193&-1.178&-0.863  &-1.254 & -0.875 & -1.014  & -0.687\\
        $ ^{177}\rm{Tl^{m}}$ & 1.963 & $(11/2^{-})\rightarrow 0^{+}$   &5\,\,  &0.012&0.137&-3.346 & -2.666 &-3.238 & -4.205 & -3.948  & -2.278\\
\hline
    \end{tabular}
    \label{table1}
\end{table*}

\begin{table}[!htb]
    \centering
    \setlength\tabcolsep{8pt}
    \renewcommand\arraystretch{1.8}
    \caption{The standard deviation $\sigma$ between the experimental proton radioactivity half-lives and calculated values obtained using Eq. (\ref{logTT}), analytical Eq. (\ref{d}) of parameter $d$, UDL, N-GNL, UFM.}
	\begin{tabular}{cccccc}	\hline
		
           {\bf Models} & {\bf Cal1}& \bf Cal2& {\bf UDL} & {\bf N-GNL} & {\bf  UFM} \\\hline
           $\sigma$ & 0.380 &0.511& 0.403 & 0.525 & 0.705 \\\hline
	\end{tabular}
        \label{table2}
 \end{table}

Furthermore, the standard deviation $\sigma$ is used to globally quantify the agreement between experimental data and calculated values. In this work, $\sigma$ is defined as
\begin{equation}
    \sigma=\sqrt{\frac{1}{n}\displaystyle\sum_{i=1}^{n} \Delta_i^2}.
    \label{sigma}
\end{equation}
The calculation results of $\sigma$ are listed in Table \ref{table2}. From the table,  it can be seen that though the absolute reduction in standard deviation $\sigma_{\rm{Cal1}}$=0.380 and $\sigma_{\rm{UDL}}$=0.403 appears modest, the $5.7\%$ relative error reduction holds significance in proton radioactivity modeling. This further indicates that our proposed improved model, which considers the effect of the centrifugal potential, is quite reliable, and that accounting for the spectroscopic factor is also necessary.

In addition, Delion derived an analytic linear relationship between the logarithm of the reduced width squared and the fragmentation potential $V_{frag}$, based on a shifted harmonic oscillator potential \cite{Delion:2009jw}. Inspired by this work, we further explore the analytic dependence of the centrifugal parameter $d$ from Eq. (\ref{d}), which connects $d$ to the oscillator frequency $\hbar \omega$, the reduced mass $\mu$, and the barrier radius $r_1$. We substitute $\hbar \omega \approx$1.5 MeV to obtain the analytically estimated $d^{\rm{Ae}}$ $\approx$ 0.167, and present the $d$ for each nucleus in the sixth column of Table \ref{table1}. Meanwhile, $V_0=56.1$ MeV was refitted using the $d^{\rm{Ae}}$, and then it was substituted into Eq. (\ref{logTT}) to obtain the proton radioactivity calculated using $d^{\rm{Ae}}$, which was expressed as ${\rm{log}}_{10}T_{1/2}^{\rm{{Cal2}}}$. As shown in Table \ref{table1} and Table \ref{table2}, the calculated half-lives using $d^{\rm{Ae}}$ exhibit larger deviations compared to the empirical fit $d=0.143$. This discrepancy may arise from simplifications in the harmonic oscillator assumption, such as neglecting nuclear deformation effects or higher-order corrections to the centrifugal potential. Additionally, the fragmentation potential $V_{frag}$ in Eq. (\ref{V_11}) depends sensitively on the choice of $r_B$, which could introduce systematic uncertainties in the analytic derivation.

Despite the theoretical appeal of Eq. (\ref{d}), the empirical parameter $d$ better captures the complex interplay between nuclear structure and tunneling dynamics, particularly for high-$l$ transitions where centrifugal effects dominate. The linear relationship between $S_{p}^{\rm{{Cal}}}$ and $V_{frag}$ in Fig. \ref{logSp-Vfrag} validates the RMF+BCS approach for spectroscopic factors, yet the decay width ultimately depends on both $S_p$ and the centrifugal-modified penetrability. Therefore, we retain the fitted $d$ in Eq. (\ref{logTT}) for final predictions, as it ensures optimal agreement with experimental data while preserving the physical interpretation of centrifugal barrier effects.

\begin{figure}[!htb]\centering
   \includegraphics
   [width=1.1\hsize]
   {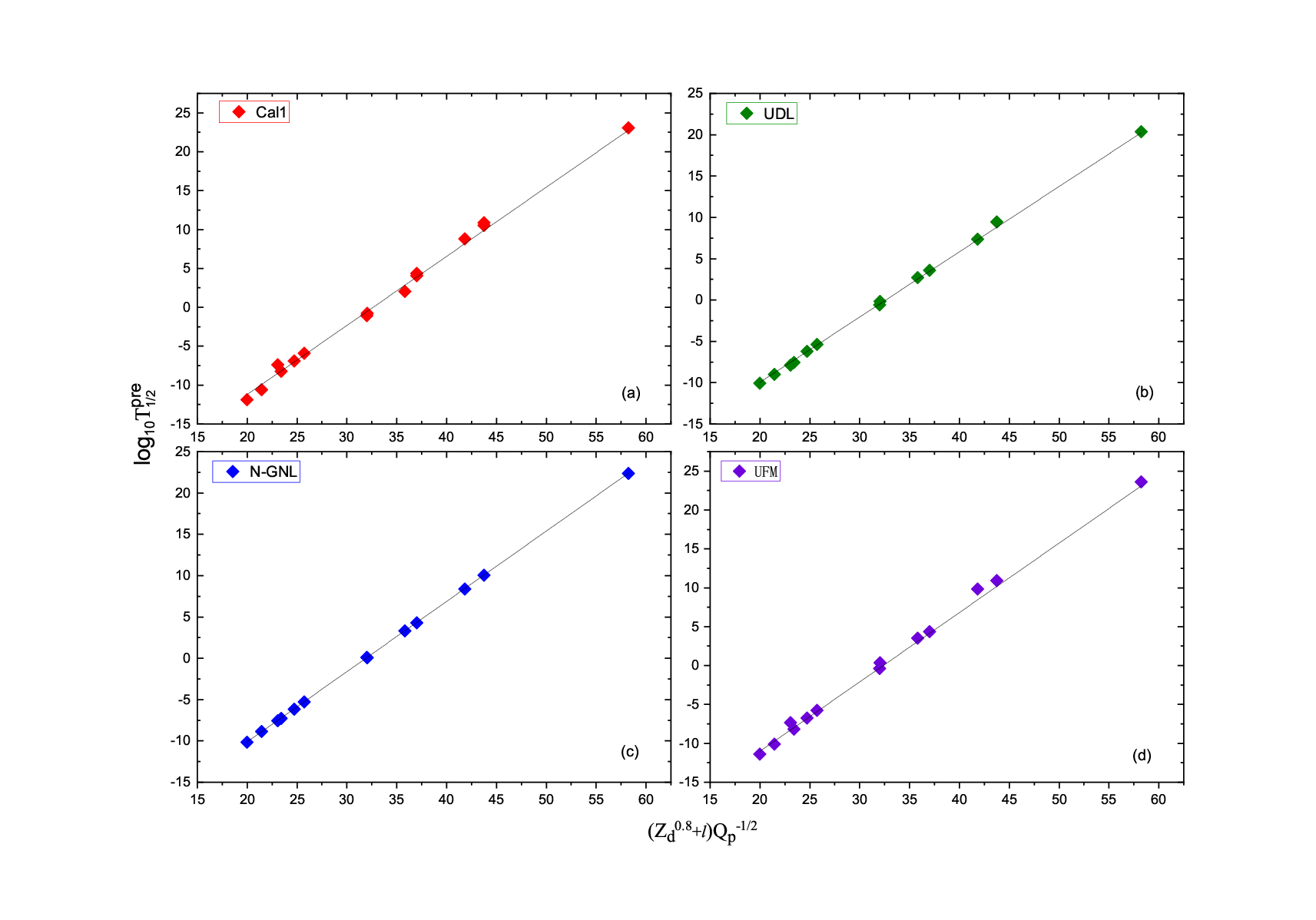}
   \caption{(color online) Relationship between the predictions of these models and/or formulas given in Table \ref{table3} and $(Z_d^{0.8}+l)+Q_p^{-1/2}$.}
   \label{predicted}
   \end{figure}

In view of the fact that the above results calculated by Eq. (\ref{logTT}) are better with fitting $d=0.143$, we further extend this model to predict proton radioactivity half-lives for possible candidates, which are energetically allowed or observed but not yet quantified in NUBASE2020. Similarly, we also use UDL \cite{Qi:2012pf}, N-GNL \cite{Chen:2019aio} and UFM \cite{jian:2010unified} for comparison. The detailed predictions are given in Table \ref{table3}, where the first three columns list the proton emitter, the released energy $Q_p$, and the angular momentum $l$. The last four columns give the predicted proton radioactivity half-lives in logarithmic form using Eq. (\ref{logTT}), UDL, N-GNL, and UFM, respectively. The table shows that the predictions from our model are relatively consistent with those of the other two models and/or formulas, particularly with the UFM. In addition, to further verify the credibility of our predictions, we plot the relationship between the logarithmic values of predicted proton radioactivity half-lives by Eq. (\ref{logTT}), UDL, N-GNL, and UFM, and $(Z_d^{0.8}+l)+Q_p^{-1/2}$ of the new Geiger-Nuttall law \cite{Chen:2019aio}. As shown in Fig. \ref{predicted} (a)-(d), all predicted proton radioactive half-lives exhibit a linear relationship with $(Z_d^{0.8}+l)+Q_p^{-1/2}$, which strongly validates the reliability of our predictions and provides substantial support for future research on the proton radioactivity half-lives of newly synthesized isotopes. 

 \begin{table}[!htb]
    \centering
    %\small
    \setlength\tabcolsep{2.5pt}
    \renewcommand\arraystretch{1.0}
    \caption{Comparison of the predicted proton radioactivity half-lives, which are observed or their proton radioactivity is energetically allowed but not yet quantified in the latest atomic mass excess NUBASE2020 \cite{Kondev:2021lzi} and the related Ref. \cite{Xu:2023etu,Zhang:2023yrn}, have been predicted using Eq. (\ref{logTT}), UDL, N-GNL, and UFM.}
    \begin{tabular}{cccccccc}	\hline
        \multicolumn{3}{c}{}  &  \multicolumn{5}{c}{$\log_{10}{T_{1/2 } (s)} $} \\\cline{5-8}
         Nuclei & $Q_{p}$(MeV)& $\,\,l\,\,$ &$\rm{S_p^{cal}}$ & Cal & UDL & N-GNL &UFM \\\hline
        $ ^{111}\rm{Cs}$ & 1.740 &2\,\,\, &0.992& -11.910 & -10.094 & -10.145& -11.406 \\
        $ ^{116}\rm{La} $& 1.591 &2\,\,\, &0.984& -10.629 & -9.000& -8.887 & -10.128 \\
        $ ^{127}\rm{Pm} $& 0.792 &2\,\,\, &0.625& -1.063 & -0.620 & 0.094 & -0.411 \\
        $ ^{137}\rm{Tb} $& 0.843 &5\,\,\, &0.961&  2.023 & 2.714 & 3.293 & 3.490 \\
        $ ^{146}\rm{Tm^{n}}$& 1.144 &5\,\,\,&0.747& -0.795 & -0.177 & 0.065 & 0.358 \\
        $ ^{159}\rm{Re}$ & 1.606 &0\,\,\,&0.745&-6.892 & -6.227  & -6.156 & -6.735 \\
        $ ^{165}\rm{Ir}$ & 1.547 &0\,\,\,&0.786& -5.924 & -5.387  & -5.303 & -5.786\\
        $ ^{169}\rm{Ir^{m}}$& 0.782 &5\,\,\,& 0.116&8.248 & 7.362& 8.363 & 9.829  \\
        $ ^{171}\rm{Ir^{m}} $&0.403 &5\,\,\, &0.152& 23.070 & 20.337& 22.337 & 23.605  \\
        $ ^{168}\rm{Au}$ & 2.007  &0\,\,\,&0.047& -7.376 & -7.887  & -7.576 & -7.326\\
        $ ^{169}\rm{Au}$ & 1.947  &0\,\,\,&0.788& -8.248 & -7.572  & -7.276 & -8.196\\
        $ ^{172}\rm{Au}$ & 0.877 &2\,\,\, &0.998& 4.059 & 3.578  & 4.284 & 4.347 \\
        $ ^{172}\rm{Au^{m}}$ & 0.627 &2\,\,\,&0.999 & 10.530 & 9.433  & 10.034 & 10.921 \\
        \hline
    \end{tabular}
    \label{table3}
\end{table}

\section{Summary}
\label{Sec.IV}
In summary, this study advances the theoretical description of proton radioactivity by refining the harmonic oscillator model to account for centrifugal barrier effects and fragmentation potential dynamics. Key results include the derivation of a linear relationship between $\log_{10}{\gamma^2}$ and $V_{frag}$, confirming the role of nuclear structure in tunneling processes. While the analytical centrifugal parameter $d^{\rm{Ae}}$ $\approx$ 0.167 from harmonic oscillator assumptions aligns broadly with experimental trends, the empirically fitted value $d$=0.143 ensures better agreement with observed half-lives, highlighting the influence of nuclear deformation and higher-order corrections. Meanwhile, the integration of RMF+BCS-derived spectroscopic factors and centrifugal corrections into the model achieves a low standard deviation $\sigma$=0.380 across 32 spherical nuclei, outperforming existing models. In addition, predictions for candidates like $^{111}\rm{Cs}$ to $^{172}\rm{Au^{m}}$ further validate the robustness of the framework through consistency with the modified Geiger-Nuttall law. This work establishes a precise tool for probing nuclear stability, particularly near proton drip lines, with future extensions anticipated for deformed nuclei and multi-proton radioactivity systems.

\section*{Acknowledgments}
Supported by the National Natural Science Foundation of China (Grants Nos: 12175100, 11975132 ).

\section*{Data Availability Statement}
The manuscript has associated data in a data repository [Authors’ comment: All data included in this manuscript are available upon request by contacting with the corresponding author.]

%\bibliography{ref}

\end{document}